\documentclass[draft]{agujournal2019}
\usepackage{url} %this package should fix any errors with URLs in refs.
\usepackage{lineno}
\usepackage{amsmath}
\usepackage[finalnew]{trackchanges}
\usepackage{soul}
\newcommand{\dlon}{\Delta lon}

\draftfalse

\journalname{Space Weather}

\begin{document}

%% ------------------------------------------------------------------------ %%
%  Title
%
% (A title should be specific, informative, and brief. Use
% abbreviations only if they are defined in the abstract. Titles that
% start with general keywords then specific terms are optimized in
% searches)
%
%% ------------------------------------------------------------------------ %%

% Example: \title{This is a test title}

\title{Prediction of Dst during solar minimum using in situ measurements at L5}
%\title{Prediction of Dst during high-speed streams using in situ measurements at L5}

%% ------------------------------------------------------------------------
%
%  AUTHORS AND AFFILIATIONS
%
%% ------------------------------------------------------------------------ %%

% Example: \authors{A. B. Author\affil{1}\thanks{Current address, Antartica}, B. C. Author\affil{2,3}, and D. E.
% Author\affil{3,4}\thanks{Also funded by Monsanto.}}

\authors{R. L. Bailey\affil{1}, C. M\"ostl\affil{1}, M. A. Reiss\affil{1,2}, A. J. Weiss\affil{1,3}, U. V. Amerstorfer\affil{1}, T. Amerstorfer\affil{1}, J. Hinterreiter\affil{1,3}, W. Magnes\affil{1} and R. Leonhardt\affil{4}}

\affiliation{1}{Space Research Institute, Austrian Academy of Sciences, Graz, Austria}
\affiliation{2}{Heliophysics Science Division, NASA Goddard Space Flight Center, Greenbelt, MD 20771, USA}
\affiliation{3}{Institute of Physics, University of Graz, Graz, Austria}
\affiliation{4}{Conrad Observatory, Zentralanstalt f\"ur Meteorologie und Geodynamik, Vienna, Austria}
% \affiliation{1}{First Affiliation}

\correspondingauthor{Rachel Bailey}{rachel.bailey@oeaw.ac.at}

% Example:
% \begin{keypoints}
% \item	List up to three key points (at least one is required)
% \item	Key Points summarize the main points and conclusions of the article
% \item	Each must be 100 characters or less with no special characters or punctuation and must be complete sentences
% \end{keypoints}

\begin{keypoints}
\item In-situ L5 data can be used for Dst forecasts at Earth and perform better than 27-day recurrence
\item Low consistency of Bz over multiple days limits the accuracy of Dst predicted from L5
\item This method performs best when forecasting the minimum Dst during SIRs
\end{keypoints}

%% ------------------------------------------------------------------------ %%
%
%  ABSTRACT and PLAIN LANGUAGE SUMMARY
%
% A good Abstract will begin with a short description of the problem
% being addressed, briefly describe the new data or analyses, then
% briefly states the main conclusion(s) and how they are supported and
% uncertainties.

% The Plain Language Summary should be written for a broad audience,
% including journalists and the science-interested public, that will not have 
% a background in your field.
%
% A Plain Language Summary is required in GRL, JGR: Planets, JGR: Biogeosciences,
% JGR: Oceans, G-Cubed, Reviews of Geophysics, and JAMES.
% see http://sharingscience.agu.org/creating-plain-language-summary/)
%
%% ------------------------------------------------------------------------ %%

%% \begin{abstract} starts the second page

\begin{abstract}
Geomagnetic storms resulting from high-speed streams can have significant negative impacts on modern infrastructure due to complex interactions between the solar wind and geomagnetic field. One measure of the extent of this effect is the Kyoto $Dst$ index. We present a method to predict $Dst$ from data measured at the Lagrange 5 (L5) point, which allows for forecasts of solar wind development 4.5 days in advance of the stream reaching the Earth. Using the STEREO-B satellite as a proxy, we map data measured near L5 to the near-Earth environment and make a prediction of the $Dst$ from this point using the Temerin-Li $Dst$ model enhanced from the original using a machine learning approach. We evaluate the method accuracy with both traditional point-to-point error measures and an event-based validation approach. The results show that predictions using L5 data outperform a 27-day solar wind persistence model in all validation measures but do not achieve a level similar to an L1 monitor. Offsets in timing and the rapidly-changing development of $B_z$ in comparison to $B_x$ and $B_y$ reduce the accuracy. Predictions of $Dst$ from L5 have an RMSE of $9$ nT, which is double the error of $4$ nT using measurements conducted near the Earth. The most useful application of L5 measurements is shown to be in predicting the minimum $Dst$ for the next four days. This method is being implemented in a real-time forecast setting using STEREO-A as an L5 proxy, and has implications for the usefulness of future L5 missions.
\end{abstract}

%\section*{Plain Language Summary}
%[ enter your Plain Language Summary here or delete this section]

% ----------------------------------------------------------------------
\section{Introduction}
% ----------------------------------------------------------------------

% State the problem
The solar wind has myriad effects on the Earth's magnetic field, among them the enhancement of the ring current around the Earth's equator \cite{Gonzalez1994}. Through streams of charged particles, the solar wind injects energy into the ring current and thereby reduces the global magnetic field strength \cite{Daglis1999}. This can have consequences on GPS and satellite communication as well as flight operations where crew may be exposed to greater levels of radiation \cite{Schrijver2015}. The extent of the enhancement and resultant reduction in field strength is often given by the \textit{disturbance storm time} index, $Dst$. This is an hourly value derived from geomagnetic field variations measured at four observatories (Honolulu, Kakioka, San Juan and Hermanus) below $35^\circ$ in geomagnetic latitude \cite{Mayaud1980}. Different levels of geomagnetic activity are often described by the minimum of $Dst$ reached, with a $Dst \leq -200$ nT denoting an extreme geomagnetic storm, $Dst \leq -100$ nT being an intense storm, and $Dst \leq -50$ nT being only a moderate storm. A $Dst \leq -30$ nT is sometimes used to denote weak geomagnetic storms. The largest negative values of $Dst$ are seen almost exclusively in storms caused by interplanetary coronal mass ejections/ICMEs \cite{Borovsky2006}, while moderate storms occur throughout the solar cycle \cite{Richardson2001, Tsurutani2006}. These more common, milder storms are driven by high-speed streams and stream interaction regions/SIRs \cite{Alves2006, Jian2006}, and \citeA{Richardson2000} showed that high-speed streams lead to 70\% of geomagnetic activity during solar minimum. In rare cases SIR-driven storms can also become major geomagnetic events \cite{Richardson2006} with an expected maximum possible storm strength of $Dst \sim -180$ nT.

% Dst prediction
Although the official Kyoto $Dst$ is derived solely from ground-based field measurements, a very good estimate of the upcoming $Dst$ can be made based on in situ solar wind data from satellites at the Lagrange 1 (L1) point \cite<e.g. most recently the DSCOVR satellite, see>{Burt2012}. This allows for a prediction lead time of 10--50 minutes, with the amount of time determined by solar wind speed between L1 and Earth. The state-of-the-art approach in this respect is the model of \citeA{Temerin2006}, an empirical technique that achieves a Pearson correlation coefficient between the observed and predicted $Dst$ values of $0.96$ for the seven years of data evaluated. This model depends solely on solar wind input, and the variation in $Dst$ at one point depends on both the solar wind at that point in time and past modelled $Dst$ timesteps.

% L5 Data
For forecasts beyond a half-hour window, we look now to possible future missions to the L5 point, which sits $60^\circ$ behind the Earth in its orbit and roughly 4.5 days in advance for corotating solar wind structures. A space weather mission at this point to perform in situ solar wind measurements has been discussed many times before \cite{Gopalswamy2011, Lavraud2016, Hapgood2017}, and presents a strong opportunity for accurate forecasts of space weather events with a much enhanced lead time ranging from hours to days. As shown in \citeA{Thomas2018}, a solar wind monitor at the L5 point can provide very good forecasts of the ambient solar wind variations, of which high-speed streams and SIRs are of primary interest. 

% What we'll do
In this work we show how predictions of the $Dst$ index at Earth can be made using L5 data and discuss the applicability and accuracy of the method. This is carried out using data from the Solar Terrestrial Relations Observatory Behind \cite<STEREO-B, NASA, see>{Kaiser2005} satellite, which crossed the L5 point in late 2009, as a proxy for a future L5 mission. Data from this satellite is mapped to L1 as if it had been measured there by correcting for both time passed in solar rotation speed and solar wind expansion \cite{Thomas2018}, and a method for predicting the $Dst$ from L1 data is then applied. The accuracy of the $Dst$ forecast is evaluated using a combination of traditional error metrics (e.g. correlation coefficient, mean error) as well as a method considering the prediction of events (e.g. $Dst$ minimum) without comparing the $Dst$ development point-to-point. Past studies have looked at predicting the general solar wind properties \cite{Simunac2009, Turner2011, Kohutova2016, Temmer2018, Owens2019} while this study aims specifically to determine how well the development of $Dst$ and geomagnetic effects can be predicted using L5 data. The results also include a brief analysis of the sensitivity of $Dst$ prediction to offsets in measurements of the magnetic field. This study serves as a verification for methods that are now being implemented in real-time using STEREO-A data as it crosses the L5 point and moves towards the Earth.

% ----------------------------------------------------------------------
\section{Methods}
% ----------------------------------------------------------------------

\subsection{Mapping L5 data to L1}

% L5 to L1 mapping
Studies using the STEREO satellites (launched in 2006) as proxies for satellites positioned at L5 have been undertaken in the past. \citeA{Simunac2009} mapped data from L5 to L1 using a time shift determined by the synodic rotation period of the Sun and showed good agreement between the solar wind speed and density measured at the two locations. \citeA{Turner2011} evaluated the correlation between time-shifted measurements ahead in the Parker spiral and L1 data. It was shown that while the correlation in solar wind speed remains high, there is rarely much correlation in magnetic field components. \citeA{Thomas2018} continued in this thread but went a step further and carried out a comprehensive analysis of solar wind forecasting skill using data measured near L5.

To map data measured at L5 or thereabouts to L1, we use the same approach as described in \citeA{Thomas2018} and apply a time shift to the data measured at STEREO-B assuming a rotation in the solar wind equivalent to the rotation speed at the solar equator of roughly 27 days ($\Delta t_{\text{lon}}$). Here we use a synodic rotation period of $T_{\text{syn}} = 27.27$ days as given in \citeA{Owens2013}. A second adjustment to the time to correct for differences in radial distances ($\Delta t_{\text{r}}$) and solar wind expansion timing is calculated as follows based on Eq. 1 from \citeA{Simunac2009}:

\begin{linenomath*}
\begin{equation}
\Delta t = \Delta t_{\text{lon}} + \Delta t_{\text{r}} = \frac{\dlon}{\Omega_{\text{Sun}}} + \frac{(r_{\text{L1}} - r_{\text{STB}})}{v_{\text{sw}}},
\end{equation}
\end{linenomath*}

where $r_{\text{L1}}$ and $r_{\text{STB}}$ are the radial distances of L1 and STEREO-B from the Sun, while $v_{\text{sw}}$ is the mean solar wind speed at the time of measurement. $\dlon$ is the difference in longitude between the Earth/L1 and STEREO-B. $\Omega_{Sun}$ is the variable for solar rotation speed $360^{\circ}/T_{\text{syn}}$. %The variable for solar rotation speed, $\Omega_{Sun} = 360/T_{syn}$ from the original equation for $\Delta t_{r}$, drops away when dividing by the rotation speed to get the shift in time.
The total time shift $\Delta t$, which varies with longitudinal distance between the satellite and Earth, is then added to the time of measurements from STEREO-B. Since this results in a new range of times with increasing difference between the new and original values as STEREO-B moves away, the measurements are interpolated back to periodic hourly time values.

A second adjustment is applied to the solar wind data to account for areal expansion of the solar wind at different radii \add{and in the Parker spiral} \cite{IntroSpacePhysics}.\add{ All variables are multiplied by a correction factor determined by the ratio between the distance of STEREO-B and L1 $(r_{\text{STB}}/r_{\text{L1}})$.} The rate of expansion for the density\remove{, total field strength, radial and vertical magnetic field components} is assumed to behave according to the inverse-square law \cite{Kumar1996}, while the \change{azimuthal field component scales as $1/r$ due to the expansion in the Parker spiral}{magnetic fielc components scale according to factors as given in} \citeA{Hanneson2020} \add{varying between -2 and -1}. \remove{All variables are multiplied by a correction factor determined by the ratio between the distance of STEREO-B and L1 $(r_{\text{STB}}/r_{\text{L1}})$ scaled accordingly by the power of -2 or -1.} In the case of STEREO-B, which was at a distance greater than $1$ AU, this means that the solar wind was corrected backwards and effectively compressed to L1.
%CM: This correction is fine for density, I am not sure about the exponent for the magnetic field (need to check papers); its definitely wrong for the solar wind speed which does not fall of with r-2 and which should be treated as constant; I guess the effect won't be much on the results but this needs to be corrected

\subsection{Prediction of $Dst$ from L1}

% Dst Prediction
There are many different models for predicting the $Dst$ index from solar wind measurements. Earlier models from \citeA{Burton1975} and \citeA{Obrien2000} achieve a reasonable level of accuracy. When using the OMNI2 data set as input and comparing the results to the true Kyoto $Dst$ values, they have correlation coefficients of $0.76$ and $0.84$ respectively. One of the most exhaustive L1-to-$Dst$ algorithms is undoubtedly the semi-empirical model developed first in \citeA{Temerin2002} for the years 1995--1999, which was later extended to the year 2002 in \citeA{Temerin2006}. This model has a linear correlation of around $0.95$ for $Dst$ in the periods the model was intended for (in good agreement with the original work), although when using it for predictions beyond this period (2002--2019), there is a linear drift away from the real $Dst$ of about $-5.3$ nT/year due to the inclusion of a time-dependent variable in the model \cite{Temerin2015}. After applying a simple linear correction for the drift, the correlation for this model in future times is still very good at 0.90 on average. Due to the dependence of the model on local time, the input for data mapped to L1 was the time it was expected to arrive there and not the time it was measured.

The \citeA{Temerin2006} method (henceforth called TL2006) makes very good predictions of $Dst$ at the Earth using either data measured at L1 or data mapped to L1 using the aforementioned time-shifting method. A small improvement can be made to the model through application of a machine learning algorithm from the Python package SciKit-Learn to provide a correction factor for the base drift-corrected TL2006 output. The algorithm is a gradient boosting regressor (GBR), which develops an ensemble of basic regressors to calculate output (correction to $Dst$) based on the provided input. The input was the same set of variables provided to the TL2006 method (solar wind speed, density, and magnetic field components along with time). Some feature engineering was applied to the input variables to provide more information to the GBR such as including a solar wind pressure term and time derivative of $B_z$ to evaluate which variables improved the $Dst$ prediction, and those that did not lead to an improvement were removed. The most important addition was the introduction a ``ring-current term'' (with both current and past values from the prior 24 hours) based on the method in \citeA{Obrien2000}. The ring-current term described most of the $Dst$ variation not accounted for by the TL2006 method, and the prediction of $Dst$ from solar wind measurements using the TL2006 method plus this GBR correction value (enhanced method or ETL2006) leads to an improved average linear correlation of 0.95 and reduced RMSE between real and predicted $Dst$ over the time range 2000--2018. In this study we apply the model using the enhancement throughout.
%CM: **not sure if a more decent description of the GBR method would be necessary (maybe a referee will ask for it); otherwise you could point the reader directly to the code by saying where it can be found**

% ----------------------------------------------------------------------
\section{Data}
% ----------------------------------------------------------------------

% OMNI/L1 Data
The NASA OMNI2 data set was used for getting the measurements of the solar wind in the near-Earth environment and the values for the Kyoto $Dst$ (the real values to which all predictive models are compared). The machine learning algorithm providing the correction to the TL2006 $Dst$ prediction method was trained on the Kyoto $Dst$ data set for 2000-2018.

% STEREO-B/L5 Data
STEREO-B one-minute resolution PLASTIC \cite{Galvin2008} and IMPACT \cite{Luhmann2008} instrument data was used as a proxy for data measured at L5. STEREO-B differs from a true L5 mission in two ways: firstly, it was constantly in motion and moving further away from the Earth in its orbit; and secondly, it was also at a greater distance from the Sun than the Earth ($r_{\text{STB}} \approx 1.05$ AU). These differences were both accounted for using the approach defined in the methods section. Beacon data, which is low-resolution data sent soon (minutes) after measurement, was used rather than the higher-quality science data that arrives later to simulate the forecasting application of this model in a real-time operation scenario. This will have an effect on the final results, although we would not expect the quality to be greatly degraded by using beacon rather than science data. The data was also downsampled via interpolation to one-hour resolution.

STEREO-B data is given in the reference frame STBHGRTN
%CM: (**which datafiles did you use for this? our STEREO-B SCEQ file? or the original STEREO-B RTN data? if its RTN these coordinates are not based on the ecliptic - we need to check this again **
, a spacecraft-centric reference frame with $x$ pointing from the Sun to the spacecraft with the Sun as the origin, $y$ as the cross-product of the rotational axis with the $x$-component, and $z$ as the normal to these two pointing out of the ecliptic. While the satellite is still in the geospace environment, the STBHGRTN coordinate system can be transformed to GSE by flipping the $x$ and $y$ directions. For $Dst$ calculation purposes, this is converted to GSM according to the algorithms given in \citeA{Hapgood1992}. The STBHGRTN frame rotates with the spacecraft as it moves away from the Earth but we assume a rotation of the solar wind with the mapping of data measured at STEREO-B to L1, and therefore can perform the same coordinate transformation to quasi-GSE (as if the measurements had been made in the geospace environment) and then to GSM. All spacecraft positions in this work are given in heliocentric Earth equatorial (HEEQ) longitudes and latitudes, in which the $z$-axis is parallel to the Sun's rotation axis and the $x$-axis is the intersection of the solar equator and solar central meridian as seen from Earth.

Because STEREO-B is always moving and only spent a short amount of time around the actual L5 point at $-60^\circ$, for the purposes of this study we consider the location of L5 to be $-60 \pm 10^\circ$ in longitude in addition to evaluating overall statistics for the range $0$ to $-110^\circ$. Comparisons of results in the next sections will refer to the two data sets as the full data set and the reduced data set. The time range in the full data set covers five years from Feb 2007
until Jan 2012.
The time range for the reduced range of angles from STEREO-B near L5 is almost six months from August 2009
%2009-08-02 
until February 2010,
%2010-01-31
which encapsulates almost the entire solar minimum, meaning this is the optimal time range to evaluate L5-based ambient solar wind prediction. 

% Persistence model
To allow comparison to a simple reference baseline model, a third data set was created as a persistence model, as this has been shown to achieve reasonable accuracy with a solar wind recurrence rate of 27.27 days \cite{Owens2013}. For this, the same OMNI2 data was taken after being shifted into the future by 27.27 days. These models are referred to as OMNI, STB and PERS in short form and plots throughout the text. In order to calculate $Dst$ reliably, any gaps in the data were linearly interpolated over.

% ICME Removal
ICMEs that occured at one location or in one data set and not the other were removed from the data using start and end times given in the HELCATS ICMECAT catalogue \cite{Moestl2017}, which covered the period of evaluation for both STEREO-B and measurements near the Earth (in this case with the WIND satellite). This was carried out for all comparison data sets after STEREO-B had been time-shifted with ICME start and end times being corrected to the new shifted times. In total, three sets of ICMEs were removed from all data sets: those at STEREO-B, those at L1, and those at L1 shifted by 27.27 days for the persistence model. This reduces the size of the total data set by 12\%.

\begin{figure}
\includegraphics[width=\textwidth]{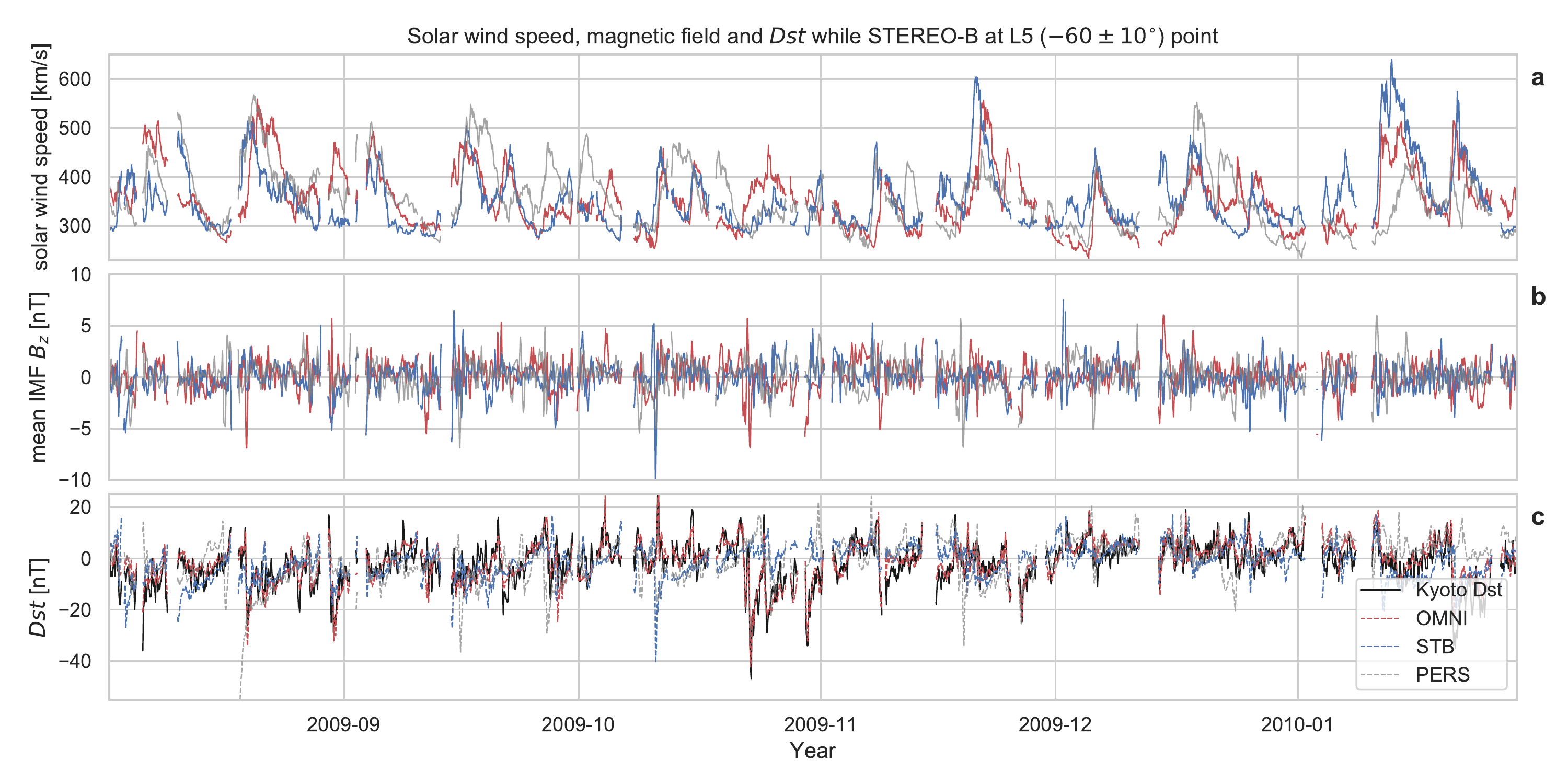}
\caption{Plot of measurements from STEREO-B (blue), the OMNI data set (red) and the persistence model data (grey) for the time period STEREO-B was closest to L5 ($-60 \pm 10^{\circ}$) in late 2009. (a) Solar wind speed from the OMNI data set and time-shifted data from STEREO-B. (b) The interplanetary magnetic field northward component, and (c) the final Kyoto $Dst$ (black) plotted against predictions made from the other data sets. Gaps in the data are ICMEs that have been removed.}
\label{fig:example}
\end{figure}

% Example plot
An example of solar wind speed, vertical magnetic field and $Dst$ values from the three data sets is plotted in Fig. \ref{fig:example}, in which the time range covered by STEREO-B when it was near the L5 point ($-60 \pm 10^{\circ}$) is shown. The STEREO-B data has already been time-shifted according to the solar rotation speed to Earth, and is plotted against the OMNI data for comparison. The high-speed streams observed at STEREO-B are easy to identify and generally overlap with those in the OMNI data, although in some cases they arrive earlier or later than their counterparts at Earth. Gaps in the data are ICMEs that have been removed. As can be seen in the comparison of $Dst$, the red line with $Dst$ predicted from the OMNI data very closely matches the actual Kyoto $Dst$, while in the mapped STEREO-B $Dst$ prediction there are still large variations. The persistence model (grey) performs worse than the STEREO-B approach.

\begin{figure}
\begin{center}
\includegraphics[width=0.7\textwidth]{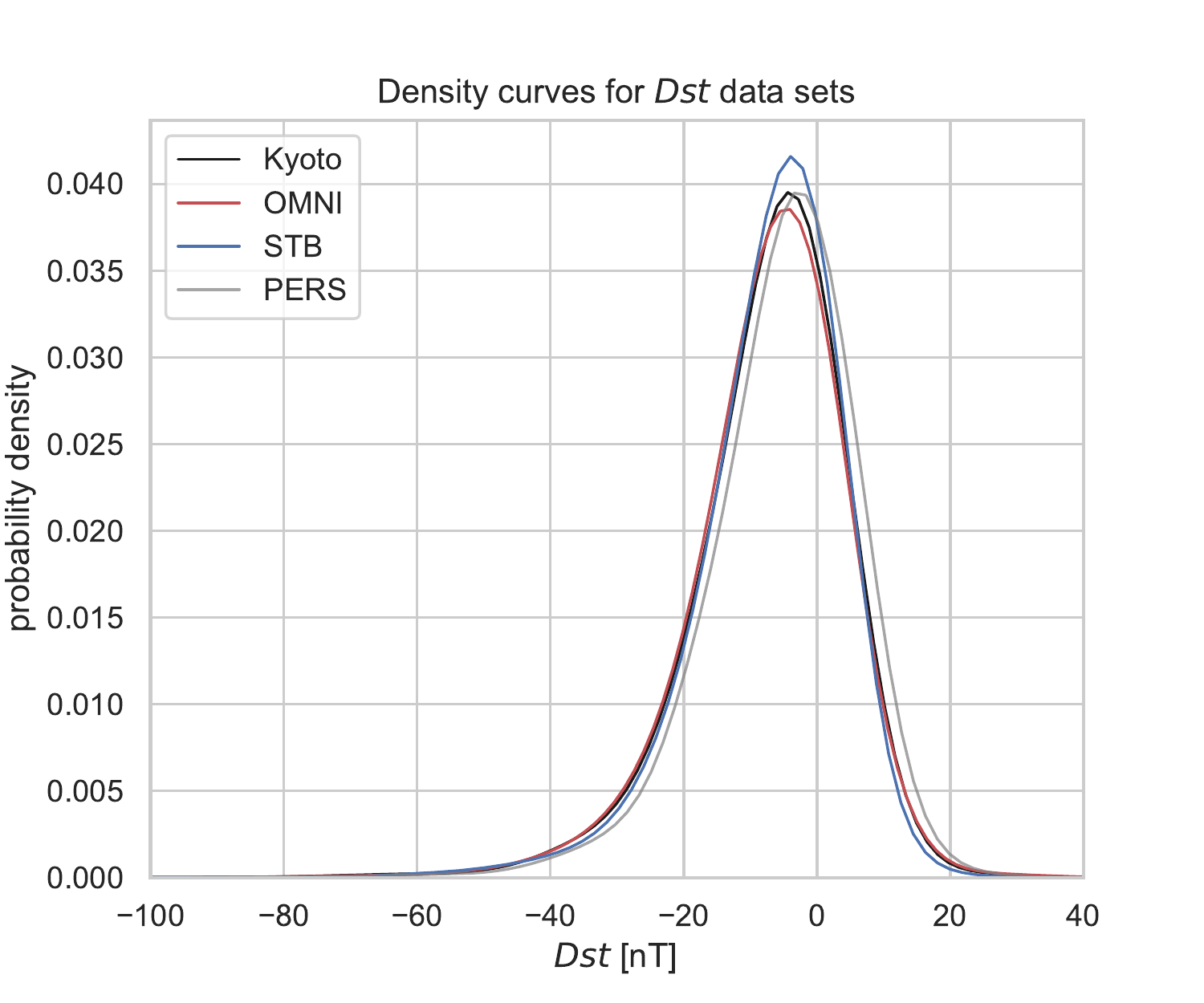}
\caption{Comparison of density curves of $Dst$ for real (Kyoto) and predicted (OMNI, STB and PERS) values with $binwidth=4^{\circ}$.}
\label{fig:density}
\end{center}
\end{figure}

The distribution of values in $Dst$ (both observed and predicted) according to the different models is plotted in Fig. \ref{fig:density}. The OMNI, Kyoto and PERS data sets have very similar distributions, but the STB $Dst$ values have a slightly higher peak close to zero and fewer values in the positive region. 
%CM: **do we really need this figure? Its a nice sanity check for ourselves but otherwise there is not much info there; the differences are negligible. **
%It is not clear why this is.

See Sec. \ref{sec:sources} for a list of all data sources.

% ----------------------------------------------------------------------
\section{Results}
% ----------------------------------------------------------------------

The goodness of prediction of $Dst$ values predicted from data measured near the L5 point will be evaluated in this section. First, the goodness will be measured according to standard error metrics, and then the models will be compared using an event-based approach looking additionally at the forecasting of events within 24-hour windows.

\subsection{Accuracy of prediction}

We first evaluate the accuracy of a prediction using data from the L5 point using standard metrics such as the Pearson correlation coefficient (PCC), mean absolute error (MAE), mean error (ME) and root-mean-square error (RMSE) of the predicted values subtracted from the observed. The results for each data set are listed in Table \ref{tab:metrics}.

Predictions from the OMNI data set consistently achieve a very good level of accuracy compared to the real Kyoto $Dst$, which is to be expected given the good behaviour of the ETL2006 model in general, although predictions using STEREO-B data are considerably worse. When comparing the STEREO-B/L5 predictions to the persistence model, we see that the prediction using STEREO-B data is better in all measures. Errors in both these models are usually twice as large as those from the OMNI model.

\begin{table*}[t]
\caption{Summary of the error measures describing the accuracy of $Dst$ predicted with various methods. The mean and minimum values are taken over the whole data set, ME is mean error, MAE is mean absolute error, RMSE is root-mean-square error and PCC is Pearson's correlation coefficient.  $Dst_{\text{all}}$ lists results for the full data set, while $Dst_{\text{reduced}}$ shows results for the data while STEREO-B was near the L5 point.}
\begin{center}
 \begin{tabular}{l c c c c c c c}
 \hline
 & & Mean & Min. & ME & MAE & RMSE & PCC \\
 \hline
 &              OMNI & -7.49 & -140.38 & -0.24 & 3.89 & 5.04 & 0.90 \\ 
 $Dst_{\text{all}}$ &  STB  & -7.28 & -172.16 & -0.21 & 8.40 & 12.25 & 0.41 \\ 
 &              PERS & -5.38 & -164.96 & -2.11 & 9.40 & 13.42 & 0.33 \\ 
 \hline
 &                  OMNI & -1.13 & -42.34 & -0.89 & 3.38 & 4.38 & 0.85 \\ 
 $Dst_{\text{reduced}}$ &  STB  & -1.45 & -40.67 & -0.56 & 6.63 & 8.89 & 0.25 \\ 
 &                  PERS & -0.83 & -62.81 & -1.19 & 8.04 & 10.90 & 0.12 \\
 \hline
\end{tabular}
\label{tab:metrics}
\end{center}
\end{table*}

\begin{figure}
\begin{center}
\includegraphics[width=0.8\textwidth]{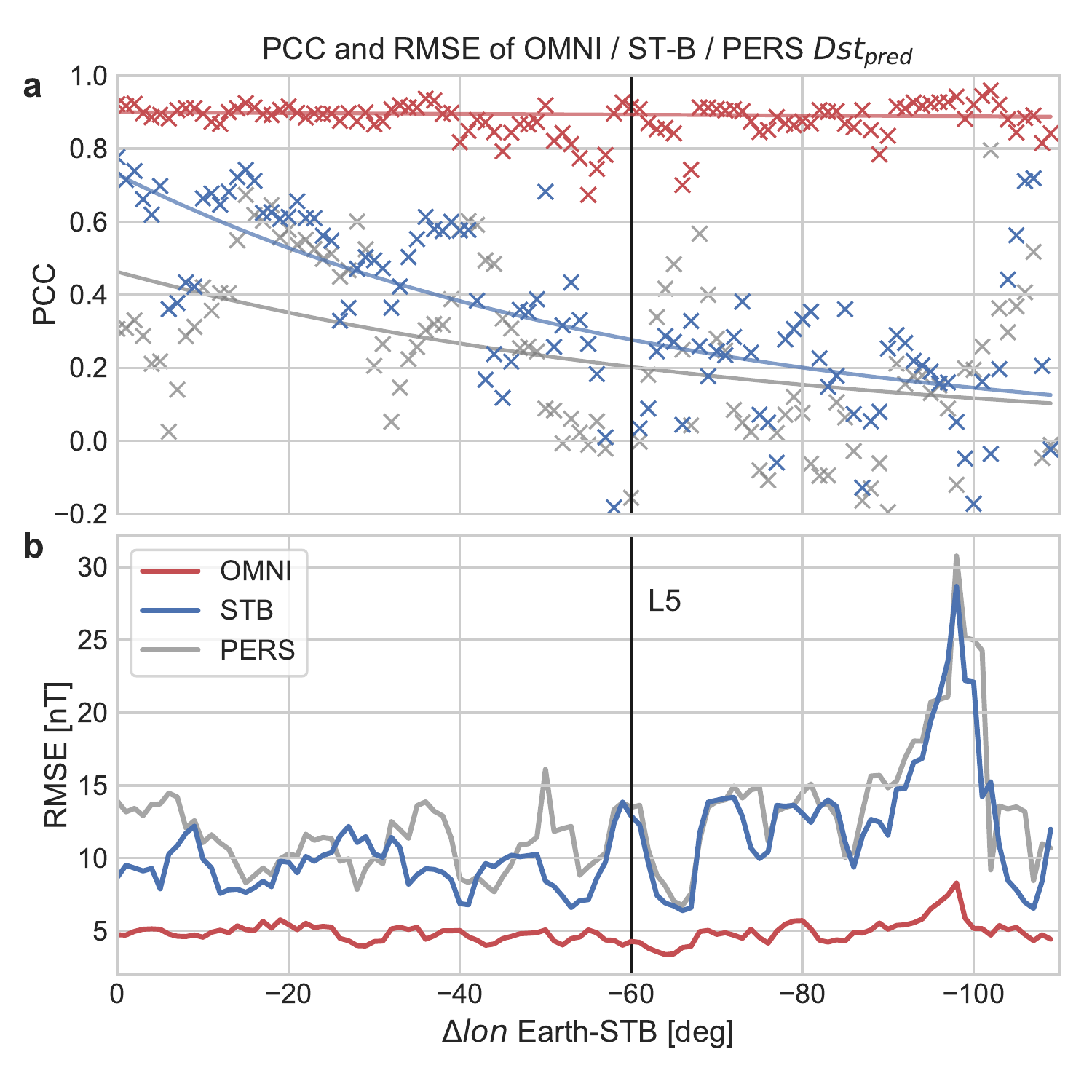}
\caption{(a) The correlation between Kyoto $Dst$ and $Dst$ predicted using OMNI (red), time-shifted STEREO-B (blue) data as a function of $\dlon$ from $0^{\circ}$ to $-110^{\circ}$, plotted against results from a persistence model (grey). (b) The RMSE for the same models over increasing $\dlon$. This range of $\dlon$ covers a range of dates from Feb 2007 till Jan 2012. The curves denote fits to the data as a decaying function of $ae^{-bx}$.}
\label{fig:corr_lon}
\end{center}
\end{figure}

In Fig. \ref{fig:corr_lon}, the PCC and RMSE are both plotted  as a function of longitudinal difference $\dlon$ between Earth and STEREO-B. Each point was calculated for the data measured at the point of longitude $\pm 3.5^{\circ}$, which led to time ranges of two to seven months being evaluated because STEREO-B was not moving away from Earth at a constant rate with regards to longitudinal distance. As can be seen in the figure, predictions of $Dst$ using OMNI data fairly consistently achieve a PCC of $0.90$ and an RMSE of roughly $5$~nT, while the accuracy of the STEREO-B prediction degrades with increasing distance from the Earth, as would be expected. Note some features of the plot at first glance seem peculiar but can be easily explained. The sudden rise in PCC beyond the $\dlon$ of -100 is a result of exceptionally quiet periods with only small values of $Dst$. The peak in RMSE just before $\dlon$ of -100 is a result of one particularly geo-effective high-speed stream ($Dst \sim -150$~nT) being observed at both STEREO-B and L1 but with an offset of multiple days.

In Fig. \ref{fig:corr_lon}a, the curves in the figure show fits to the PCC values as a decaying function of $ae^{-bx}$, and we see that at a $\dlon$ of $-110^{\circ}$, the predictions from STEREO-B data seem to have dropped to roughly the same accuracy as the persistence model. The PCC here starts at $0.80$ close to the Earth and drops to 0 at around $\dlon = -110^{\circ}$. 

\begin{figure}
\begin{center}
\hspace*{-1cm}
\includegraphics[width=1.2\textwidth]{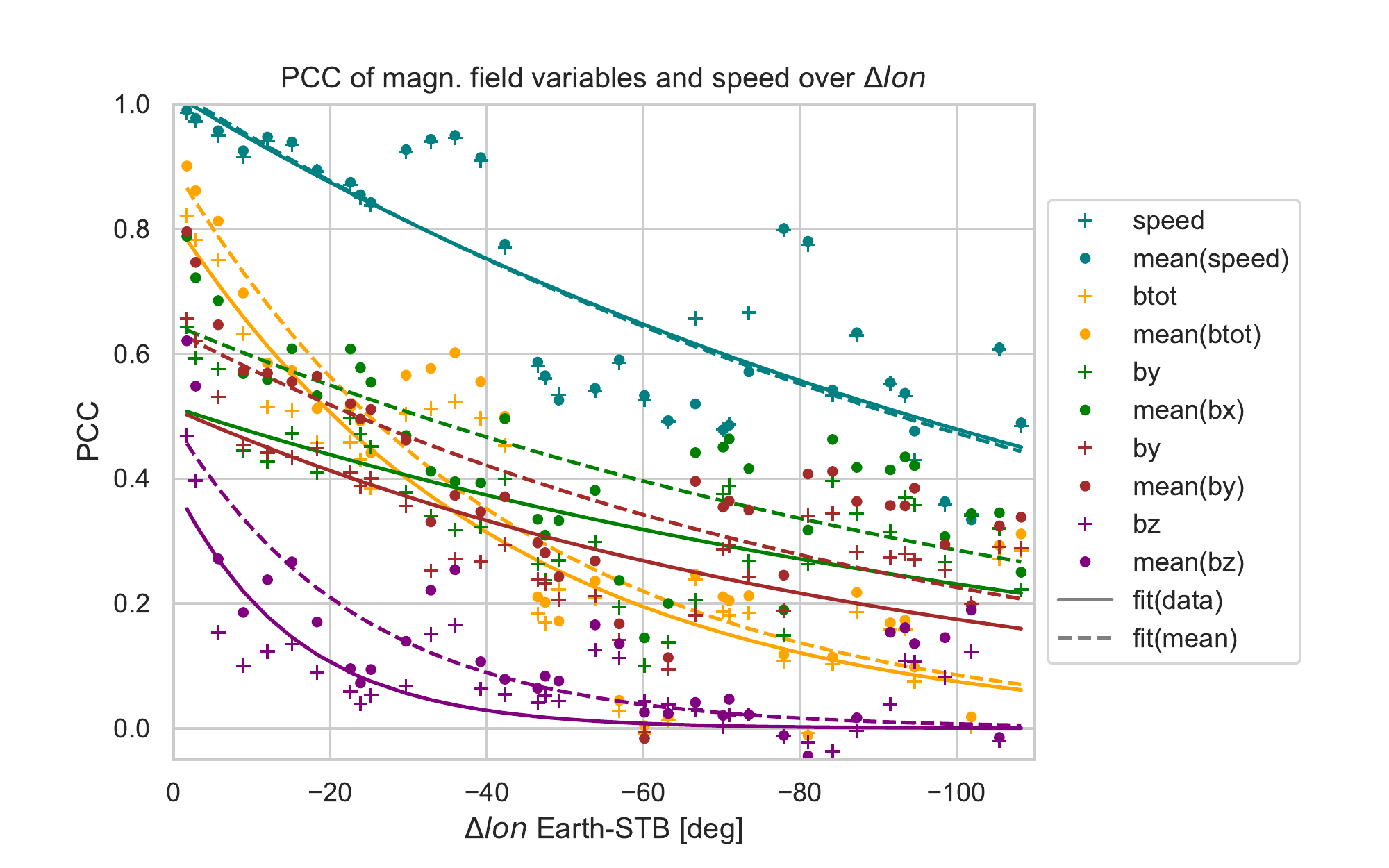}
\caption{Correlation between solar wind magnetic field measurements at L1 and STEREO-B as a function of increasing $\dlon$. '+' markers show the correlations for the original data, 'x' markers correlations for 4-hour means. The curves denote fits to the data as a decaying function of $ae^{-bx}$. Solid curves are fits to the original 1-hour data, dashed curves are fits to 4-hour means of the data.}
\label{fig:corr_vars}
\end{center}
\end{figure}

Fig. \ref{fig:corr_vars} explains this quick drop-off showing the correlations between solar wind variables in OMNI and STEREO-B with increasing $\dlon$, both as hourly values and the means over a window of 4 hours. The solar wind speed remains highly correlated, which is to be expected due to the rotation of coronal holes and the relatively slow temporal development of high-speed streams. The correlation for total magnetic field and the $x$- and $y$-components also remains good throughout, but the $z$-component drops off quickly and the correlation has already nearly reached 0 at $\dlon = -10^{\circ}$. Due to the strong dependence of $Dst$ on $B_z$ \cite<see e.g.>{Gonzalez1987}, it is not surprising that we also lose accuracy in $Dst$ prediction fairly quickly. What is however useful to note from Fig. \ref{fig:corr_vars} is that the correlation in mean($B_z$) and the other components remains good beyond the point at which the correlation of hourly values has dropped significantly.

In Fig. \ref{fig:corr_lon}, the results from the persistence model are also included in grey, and we see that there is a similar downward slope and decreasing correlation over time even though we would expect this to stay constant. This suggests that, for this time period at least, there may have been other effects in the solar wind leading to a reduction in accuracy from the prediction due to more rapid changes in the solar wind structures and high-speed streams. This is likely explained by the increase in solar activity as the Sun entered the rising phase of the solar cycle in 2011/2012. We would observe greater numbers of short-duration coronal holes closer to solar maximum in addition to the more regular coronal holes and high-speed streams that dominate the variations otherwise. Although time does not increase linearly from point to point in Fig. \ref{fig:corr_lon}, on the whole it spans five years and we can look at the general development of PCC and RMSE over time. Both show a gradual decrease in accuracy as solar activity increases, which is most important to note for the persistence model that serves as a benchmark for the others. When taking this into account, a prediction using STEREO-B data at greater values of $\dlon$ may also be a reasonable approach that is simply not represented well here due to the more active period evaluated.

%\begin{figure}
%\begin{center}
%\includegraphics[width=0.5\textwidth]{corr_over_crot.pdf}
%\caption{Error measures PCC (a) and RMSE (b) between real and predicted $Dst$ as a function of Carrington rotation. The lines have been smoothed over a window with a length of 11 rotations.}
%\label{fig:corr_over_crot}
%\end{center}
%\end{figure}

At the L5 point, the PCC for the STEREO-B model is around $0.30$. Unfortunately, looking at the points alone we see that the approach performed particularly badly at precisely this angle, although from the overall statistics we can deduce that this was only a short-duration dissonance that had little to do with the spacecraft position itself. This may instead be a result of latitudinal difference between the two satellites, which we look at in the following.

\begin{figure}
\hspace*{-2.4cm}
\includegraphics[width=1.3\textwidth]{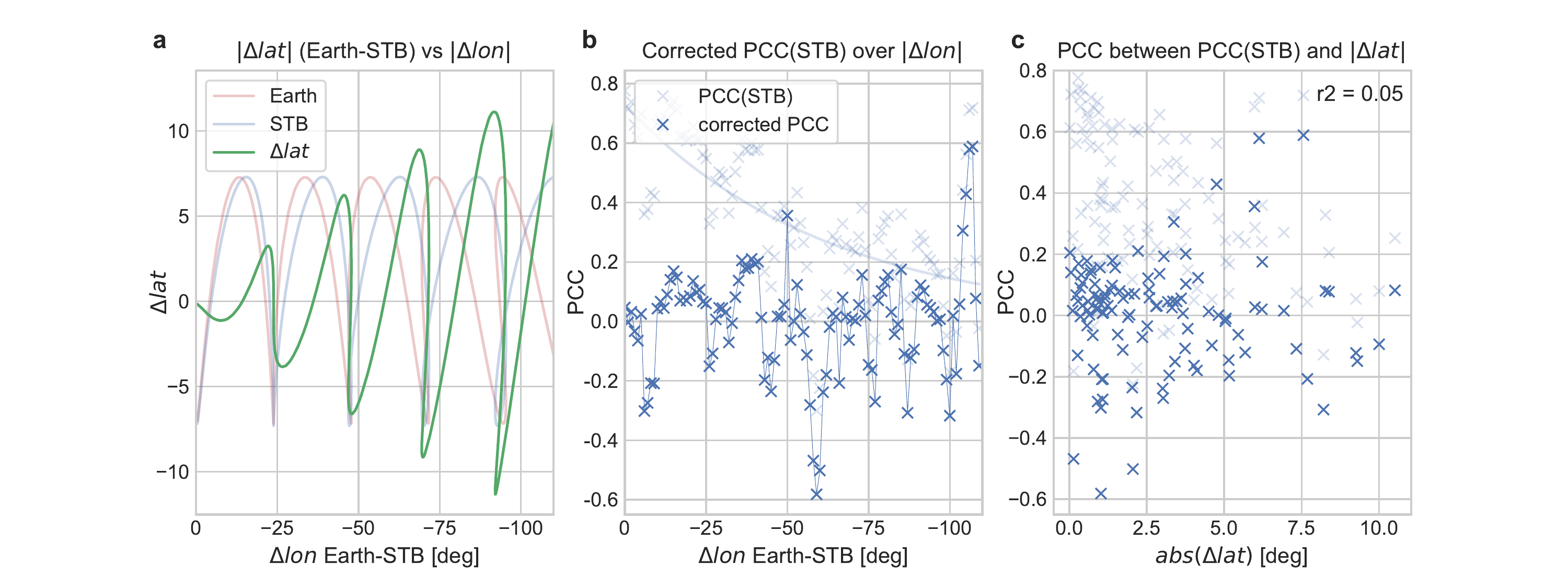}
\caption{(a) Latitudinal difference $\Delta lat$ (green) between STEREO-B (blue) and Earth (red) as a function of $\dlon$. (b) Mean error between prediction using STEREO-B data and Kyoto $Dst$, corrected for longitudinal effects. (c) Corrected PCC plotted as a function of absolute latitudinal difference $\Delta lat$.}
\label{fig:corr_lat}
\end{figure}

The dynamics of the outflowing solar wind in the heliospheric current sheet have a strong latitudinal dependence, with differences of a few degrees resulting in large variations in the shape and timing of the arriving solar wind structures. In \citeA{Simunac2009} and \citeA{Thomas2018}, for example, differences in latitude between two spacecraft were shown to have a notable effect on forecasting skill. To quantify this effect in this study, we look at the accuracy metrics with the longitudinal dependence removed plotted against the absolute difference in latitude between the two measuring points, STEREO-B and Earth. See Fig.~\ref{fig:corr_lat} for a depiction of this approach. This was achieved by subtracting the slope of the longitudinal dependence for STEREO-B (centre plot, light blue line) along with the slope in the OMNI values to account for unrelated changes over time. Interestingly, few of the metrics show any correlation with a difference in latitude, with the exception being the MAE and RMSE, which correlate mildly with $\Delta lat$ with a PCC of $-0.35$ (increasing accuracy with increasing $\Delta lat$, a confusing result). All other metrics have correlations $<$ $0.15$, such as the PCC showed as an example in the figure (centre and right). If the range of $\dlon$ is reduced to $0$ to $-40$ (in which the most periodic rotational behaviour was observed), a more predictable set of behaviour emerges. The correlation values for PCC, RMSE and MAE with $\Delta lat$ are -0.21, 0.24 and 0.15 respectively, suggesting small dependencies on $\Delta lat$ leading to a decrease in accuracy. Since an analysis of the whole range did not show the same pattern, we can deduce that discrepancies caused by the increase in solar activity later in the cycle far outweigh the effects of latitudinal difference in regards to accuracy.

As a sanity check, the same metrics with and without ICMEs were also compared, and the results were as expected. Predictions of $Dst$ using OMNI data were equally good with and without ICMEs, but there was a small improvement in the metrics for data predicted using the STEREO-B data and the persistence model because the transient events unrelated to their measurements had been removed. An analysis of errors with the input solar wind data split into high-speed streams and slow solar wind was also carried out, with the results showing that the errors in $Dst$ calculated for fast solar wind (above a threshold 1.25 times the median speed or $\sim 480$ km/s) are slightly ($\sim 20-30$ \%) larger than those for slow solar wind. \add{In a further test, we also looked at the same evaluation without scaling of the magnetic field components to correct for different orbital distances to see if this step was necessary. Not scaling the solar wind input led to a very small increase in the errors (e.g. an RMSE of $12.25$ nT increasing to $12.83$ nT), showing that the scaling of magnetic field to correct for a distance of 0.1 AU (at maximum) does have an effect on the accuracy on the forecast, although it is minimal.}

%Fast Solar Wind:
%0.46 4.62 5.98 -16.97 11.93 -17.43 12.19 0.88
%-0.56 10.27 14.89 -16.97 11.93 -16.41 12.19 0.24
%-5.88 11.33 15.62 -16.97 11.93 -11.09 11.56 0.24
%-0.00 8.70 11.93 -16.97 11.93 -16.97 0.00 0.00

%Slow Solar Wind:
%-0.45 3.68 4.73 -4.74 9.83 -4.30 9.38 0.88
%0.97 7.95 11.34 -4.74 9.83 -5.71 9.30 0.30
%-1.02 8.84 12.71 -4.74 9.83 -3.72 10.67 0.24
%0.00 7.21 9.83 -4.74 9.83 -4.74 0.00 0.00

\subsection{Event-based analysis}

An interesting alternative to standard point-to-point measures that quantitatively assess the magnitude of the forecast error at each time step is to consider each time step as an event/non-event. The primary advantages of an event-based validation analysis are as follows. Firstly, periods of weak, moderate and enhanced geomagnetic activity are weighted equally by simple point-to-point comparison measures. However, end-users of operational real-time space weather models are usually more interested in accurately predicting times of enhanced $Dst$ values, while the exact evolution in time of the $Dst$ is in most cases of secondary importance. Secondly, outliers in the predicted times can have a significant influence on the error measures and correlation coefficients determined. In context, an efficient approach is to label each time step in the predicted and observed $Dst$ timeline as an event/non-event \cite{Owens2018}. An example of this approach applied to high-speed streams can be found in \citeA{Reiss2016}, in which the OSEA software used for this analysis was also used.

In this study, we define an ``event'' as any time step the $Dst$ exceeded (or went below) a certain threshold. In this case we defined $-35$ nT as the threshold for a weak geomagnetic storm. The value of $-50$ nT, usually the threshold for a moderate storm, was also seen in the period evaluated, but did not occur often enough to allow for a reasonable statistical analysis. Similarly, $Dst$ values beyond $-100$ nT arose so infrequently that it was not worth considering them here, and we restrict ourselves to looking at all events below a level of $-35$ nT as a proxy for ``mild geomagnetic activity''. Using the corresponding threshold values, we label each time step as an event or non-event in the measured and predicted time series, and for the case of $Dst$ everything below the threshold was an event. By cross-checking the events/non-events between true and predicted $Dst$, we count the number of hits (true positives; TPs), false alarms (false positives; FPs), misses (false negatives, FNs) and correct rejections (true negatives; TNs) and summarise them in the so-called contingency table. From the entries of the contingency table, we can compute different skill measures, including the True Positive Rate TPR = TP/(TP + FN) and the False Positive Rate FPR = FP/(FP + TN). While the TPR is the proportion of correctly predicted events among all the events, the FPR is the proportion of non-events wrongly predicted as events.
%\footnote{For the reader who is more used to terms 'recall' and 'precision', we note that recall corresponds to the TPR, and precision indicates the proportion of correctly predicted events among all predicted events (FPR - 1).}

Moreover, we compute the Threat Score TS = TP/(TP+FP+FN) as a measure of the model performance, Bias B = (TP+FP)/(TP+FN) indicating whether the number of observations is underforecast ($B < 1$) or overforecast ($B > 1$), and the True Skill Statistics TSS = TPR - FPR as a measure of the overall model performance. The TSS is defined in the range [-1,1] where a perfect prediction would be equal to 1 (or -1, for a perfect inverse prediction), and a TSS equal to 0 indicates no predictive ability of the forecast model. It is important to note that the TSS is unbiased by the proportion of predicted and observed events \cite{Hanssen1965, Bloomfield2012}. Since the number of non-events exceeds the number of events by 10 to 1, the TSS is very well suited for the validation analysis conducted in this study. For a more thorough discussion of the skill measures applied here, we would like to refer the interested reader to \citeA{Jolliffe2003}. 

\begin{table*}[t]
\caption{Statistical comparison of the number of hits (true positives; TPs), false alarms (false positives; FPs), misses (false negatives; FNs), and correct rejections (true negatives; TNs), together with the True Skill Statistics (TSS), Bias and the area under the curve (AUC) for different prediction models. (A comparable AUC for the 24h minimum $Dst$ events could not be calculated due to a difference in the distribution and definition of event threshold.)}
\begin{center}
 \begin{tabular}{l c c c c c c c c c}
 \hline
 & & $N_{events}$ & TP & FP & FN & TN & TSS & Bias & AUC \\
 \hline
 &              OMNI & 843 & 618 & 280 & 225 & 36622 & 0.73 & 1.07 & 0.94 \\ 
 $Dst_{\text{all}}$ &  STB & 843 & 127 & 729 & 716 & 36173 & 0.13 & 1.02 & 0.74 \\ 
 &              PERS & 843 & 75 & 581 & 768 & 36321 & 0.05 & 0.78 & 0.70 \\ 
 \hline
 &                  OMNI & 6 & 4 & 0 & 2 & 3701 & 0.67 & 0.67 & 0.92 \\ 
 $Dst_{\text{reduced}}$ &  STB & 6 & 0 & 1 & 6 & 3700 & 0.00 & 0.17 & 0.63 \\ 
 &                  PERS & 6 & 0 & 17 & 6 & 3684 & 0.00 & 2.83 & 0.59 \\ 
 \hline
 &                  OMNI & 218 & 177 & 40 & 41 & 2886 & 0.80 & 1.00 & - \\ 
 $Dst_{\text{all}}(24h Min.)$ & STB & 218 & 65 & 131 & 153 & 2795 & 0.25 & 0.90 & - \\ 
 &                  PERS & 218 & 43 & 117 & 175 & 2809 & 0.16 & 0.73 & - \\ 
 \hline
\end{tabular}
\label{tab:events}
\end{center}
\end{table*}

Table \ref{tab:events} shows the contingency table entries and the skill measures computed over the full and reduced time ranges. We find that the prediction using OMNI/L1 data has a very high level of accuracy in all measures, and in comparison the prediction from STEREO-B/L5 is only somewhat better than a persistence model. Both STB and PERS models show a tendency to underforecast when considering the full data set.

Interestingly, we see that for the time STEREO-B was around L5 (in the reduced data set), it managed to forecast 0 of the events below the threshold, while the persistence model also forecast 0 but achieved an impressive number of FPs. It is hard to carry out statistics for this period because it was an extremely quiet time, with only 6 events below the $-35$ nT threshold over six months. To look at a less time-sensitive approach, we consider an additional forecasting method.

The bottom rows of Table \ref{tab:events} show the event-based analysis for the prediction of the minimum $Dst$ in the next $24$ hours (with a $12$-hour resolution), and in this way we consider the predictions while ignoring possible errors in timing of $\pm 12$ hours. (For the OMNI model, with a maximum forecast time of 30--60 minutes, this is obviously a pointless measure, but it is left in for comparison.) Here it becomes clear that the L5 monitor outperforms the persistence model while also achieving a higher TSS score than when simply applied to all data. In this case, the ratio of FPs to TPs is also greatly reduced from around 6 to 2.

A straightforward approach to illustrate the trade-off between the proportion of correctly predicted events (TPR) and the proportion of erroneously predicted events (FPR) for different event thresholds is the so-called receiver operator characteristic (ROC) curve. ROC curves are a helpful diagnostic to compare and quantitatively assess the predictive skill of forecast models. They illustrate how the number of correctly classified events varies with the number of incorrectly classified non-events for each model investigated here. In other words, they show the trade-off between the completeness of events and the contamination with non-events. To illustrate the predictive abilities of the different forecast models in a single summary variable, we also compute the area under the curve (AUC) defined between 0 and 1, where the best results are equal to 1.

\begin{figure}
\begin{center}
%\hspace*{-3cm}
\includegraphics[width=0.8\textwidth]{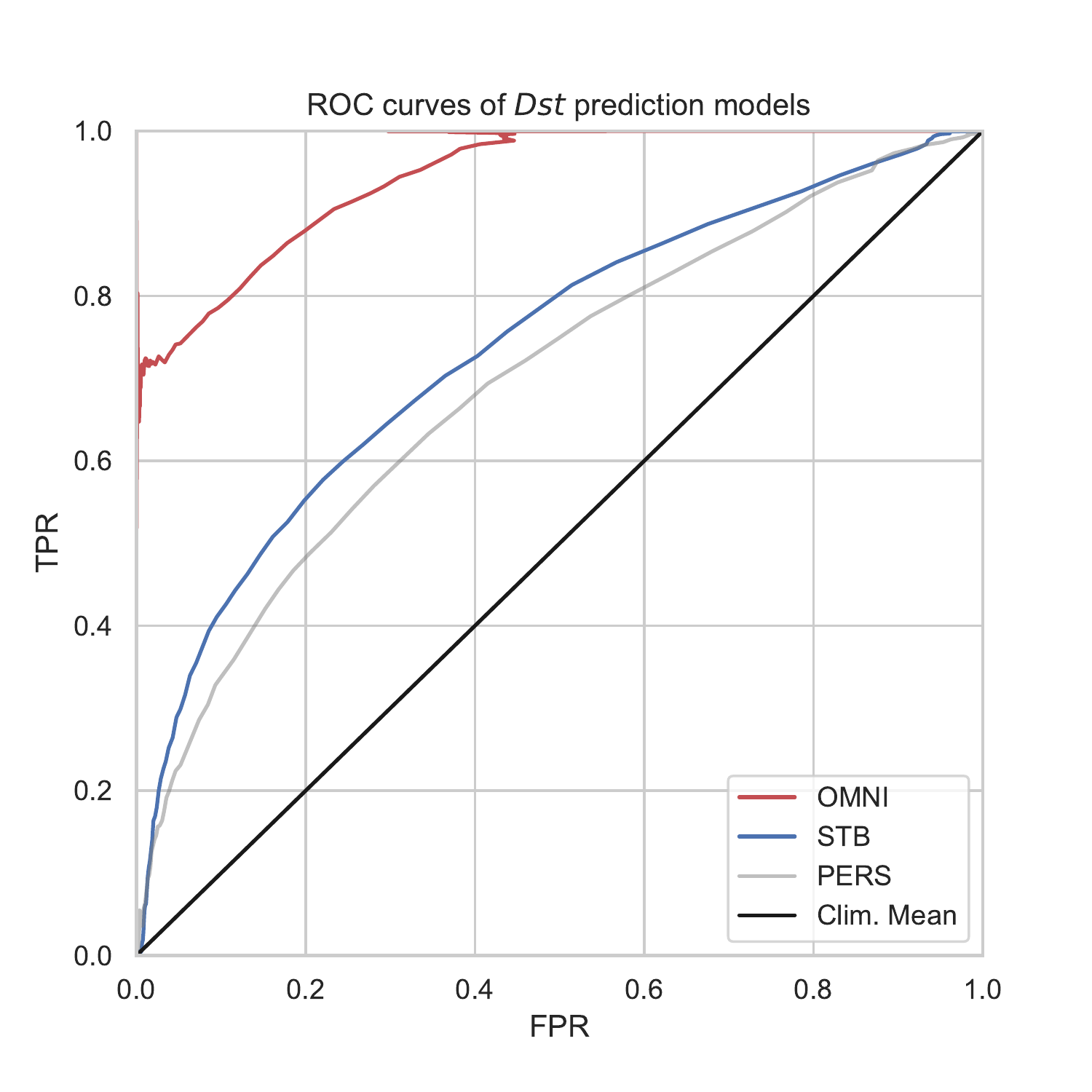}
\caption{ROC curves for all three $Dst$ prediction models (OMNI: red, STB: blue, PERS: grey) considered. The black line denotes the climatological mean (simply using the mean $Dst$ as a prediction) as baseline comparison.}
\label{fig:roc}
\end{center}
\end{figure}

Fig. \ref{fig:roc} shows the ROC curves for all models using the full data sets. At a glance it is clear that the OMNI model far outperforms the models using STEREO-B and PERS data, and the STEREO-B model is almost always somewhat better than persistence.

\subsection{Sensitivity of $Dst$ model to field measurements}

\begin{figure}
\begin{center}
\includegraphics[width=0.8\textwidth]{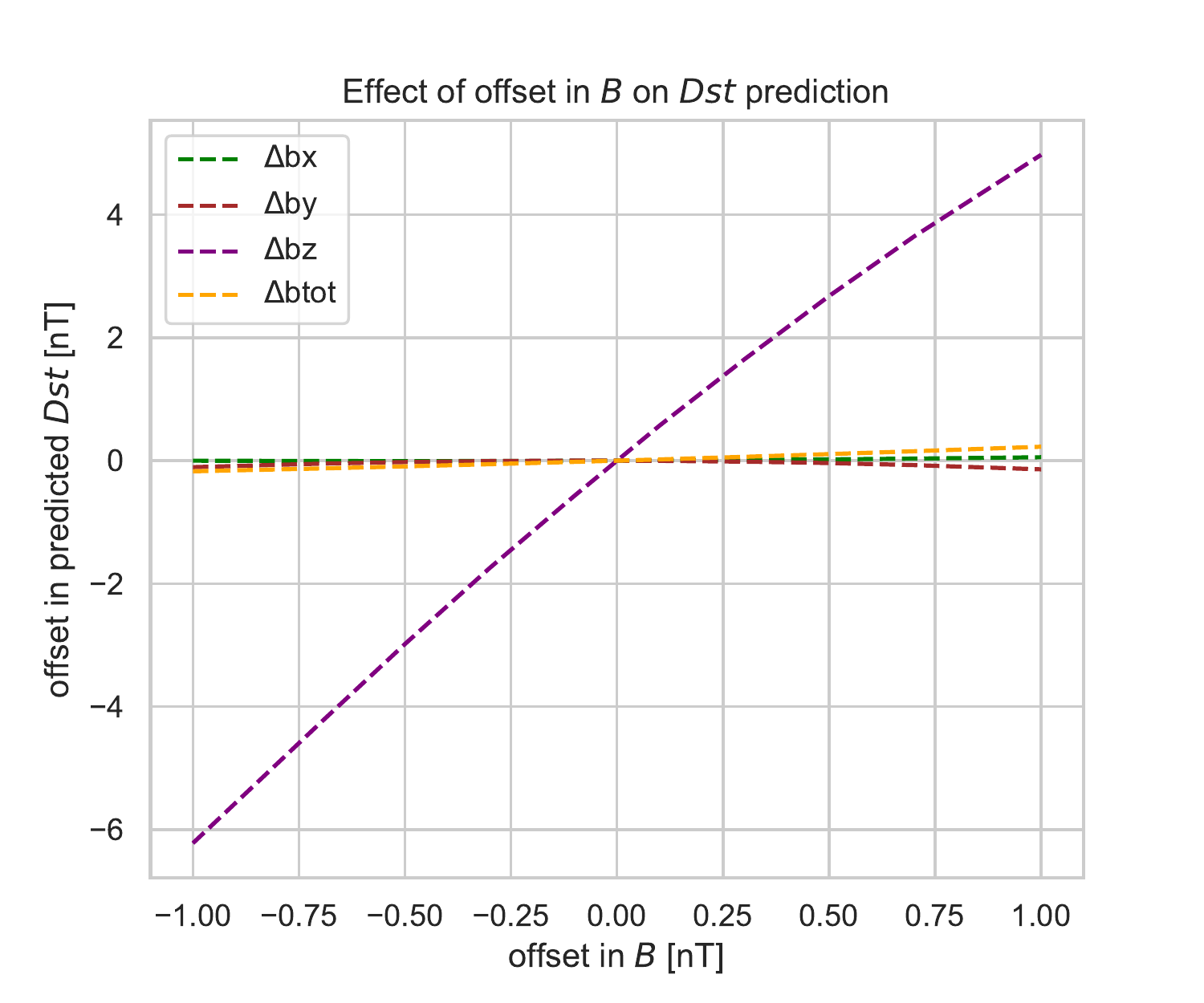}
\caption{Plot of the offset in predicted $Dst$ as a function of offset in magnetic field components (all). Only offsets in $B_z$ (purple) show a notable effect on the value of predicted $Dst$.}
\label{fig:dst_offset}
\end{center}
\end{figure}

Here we also include a brief evaluation of the sensitivity of the ETL2006 prediction model to magnetic field measurements and the effect of possible offsets in the measurements. To achieve this, the ETL2006 model was used to predict $Dst$ from two years of OMNI data to simply assess model sensitivity independent of any other factors. Fig. \ref{fig:dst_offset} shows the dependence of predicted $Dst$ on offsets in magnetic field measurements. An error of $\pm 1$ nT in the $B_z$ component leads to an error of $+5/-6$ nT in the predicted $Dst$, although offsets in all other components have almost no effect. This is useful to know for future L1 and L5 missions, which rely on in-flight calibration methods such as that described in \citeA{Plaschke2019}, to have an estimate of the error in the $Dst$ prediction if the error in the magnetic field measurements are known.

% ----------------------------------------------------------------------
\section{Discussion and conclusions}
% ----------------------------------------------------------------------

In order to evaluate the possibility of forecasting the $Dst$ index from the L5 point with a lead time of a few days, we have looked at in situ solar wind data measured by STEREO-B as it crossed the L5 point. This was mapped to L1 in time and space according to the solar rotation speed and expansion of the solar wind and high speed streams. The mapped data is used to make a prediction of $Dst$ and the results were then compared to the Kyoto $Dst$ as well as $Dst$ predicted from an L1 monitor and a 27-day persistence model. This method is useful for predicting geomagnetic effects from high-speed streams and SIRs, but can not provide forecasts for transient events such as ICMEs. The results can be summarised as follows:
\begin{itemize}
    \item A prediction of the $Dst$ index from data measured at L5 does not achieve a level of accuracy similar to predictions made using L1 data, but it performs better than a 27-day solar wind persistence model in all standard measures. Geomagnetic effects in particular are hard to predict due to the rapidly-changing development of $B_{z}$.
    \item The error in the prediction can be quantified using the MAE, which at the L5 point in the STB model has an average value of 8 nT, which is double the error from L1 at 4 nT. Offsets in $B_z$ measurements of $\pm 1$ nT can cause an offset error in the predicted $Dst$ of $+5/-6$ nT.
    \item As the values of magnetic field variations correlate much more strongly when looking at mean values, a method of predicting the $Dst$ from means taken over multiple hours would likely be a strong forecasting method, and this should be considered for operational purposes.
    \item The usefulness of L5 data in predicting $Dst$ minima in the next 24 hours is reasonable and performs better than a persistence model.
    \item A strong dependence of the accuracy of the predicted $Dst$ on latitudinal difference between the L5 proxy and L1 measurements could not be determined.
\end{itemize}

Unsurprisingly, because $B_z$ is so badly predicted from L5 due to the rapid changes in magnetic field \cite<as discussed in detail in>{Thomas2018}, it is difficult to quantify the geo-effectiveness of a high-speed stream in advance using data from somewhere as far-removed as L5 because of the strong dependence on $B_z$ in causing geomagnetic effects \cite{Gonzalez1987}. This was compounded by the fact that the coherence of high-speed streams when STEREO-B was closest to the L5 point was uncharacteristically low, and so most of the results have been based on general trends in forecasting capabilities across a much wider range of longitudes from the Earth from 0 to $-110^{\circ}$, at which point the error appeared to match that seen in the persistence model. As described in \citeA{Verbanac2011}, $Dst$ also has a strong correlation with the solar wind speed, which shows that $B_z$ is not the only input to the model providing valuable information on how the $Dst$ develops. A further effect that may lead to reduced accuracy is stream-stream or stream-CME interaction. In some cases, a CME bracketed in an SIR can give rise to enhanced geo-effectiveness in the stream \cite{Chen2019}. This would be particularly effective if such a case were to occur at L5, drastically changing the nature of the predicted high-speed stream that would later arrive at Earth. Regardless, we have shown that predictions of $Dst$ from data at L5 certainly outperform a persistence model, especially when looking at predictions of $Dst$ minimum in the next 24 hours or days.

\add{This work was carried out to validate application of the methods in an operational setting, and as such STEREO beacon data, which arrived in near real-time, was used. Beacon data is, however, of considerably lower quality than the science data, which arrives much later. Due to developments in on-board processing since the STEREO launch in 2006, real-time data to be expected from upcoming space weather satellites should be of better quality than the original beacon data and may be closer in quality to the science data. For comparison, the same parameters as discussed above were evaluated for STEREO Level 2 scientific data, and an increase in accuracy was observed. This was on a small scale (a correlation of $0.43$ instead of $0.41$, for example), but shows that the data from newer satellites with more advanced real-time data transfer should achieve a slightly higher accuracy for $Dst$ predictions than presented here.}

At the moment, STEREO-A is slowly approaching the Earth in its orbit, meaning that it can function as a proxy for real-time L5 forecasts for both the solar wind variables and $Dst$. This study has functioned as a verification for a real-time model currently running using STEREO-A data, which at the time of writing is at a longitude of $-80^{\circ}$. For this forecast, we can assume an error in $Dst$ predicted of $\pm 8$ nT from the average MAE at the L5 point reached by STEREO-B. The knowledge gathered throughout this real-time forecasting using STEREO-A will be invaluable in preparation for setting up effective predictive methods relying on a real future L5 space weather mission.

% ----------------------------------------------------------------------
\section{Sources of Data and Supplementary Material}\label{sec:sources}
% ----------------------------------------------------------------------

\noindent Solar wind in situ data:

OMNI2: \url{https://spdf.gsfc.nasa.gov/pub/data/omni/low_res_omni/}
\newline STEREO-B (BEACON): \url{https://stereo-ssc.nascom.nasa.gov/data/beacon/behind/}
\newline STEREO-B (L2): \url{https://stereo-ssc.nascom.nasa.gov/data/ins_data/impact/level2/behind/magplasma/}

\noindent Catalogues:

ICMECAT: \url{https://doi.org/10.6084/m9.figshare.4588315.v1}

\noindent Software: 

Jupyter Notebook for this work: \url{https://doi.org/10.6084/m9.figshare.11733909}
\newline Predstorm (Python, data handling): \url{https://doi.org/10.5281/zenodo.3750749}
\newline OSEA (Matlab, statistical analysis): \url{https://doi.org/10.5281/zenodo.3753104}
\newline HelioSat (Python, data download): \url{https://doi.org/10.5281/zenodo.3749561}

%%%%%%%%%%%%%%%%%%%%%%%%%%%%%%%%%%%%%%%%%%%%%%%%%%%%%%%%%%%%%%%%
%
%  ACKNOWLEDGMENTS
%
% The acknowledgments must list:
%
% >>>>	A statement that indicates to the reader where the data
% 	supporting the conclusions can be obtained (for example, in the
% 	references, tables, supporting information, and other databases).
%
% 	All funding sources related to this work from all authors
%
% 	Any real or perceived financial conflicts of interests for any
%	author
%
% 	Other affiliations for any author that may be perceived as
% 	having a conflict of interest with respect to the results of this
% 	paper.
%
%
% It is also the appropriate place to thank colleagues and other contributors.
% AGU does not normally allow dedications.

\acknowledgments

The Python packages for data download and handling that this work is based on (Predstorm v0.1, HelioSat v0.3.1) are open-source and under development and are available on our group GitHub page (\url{https://github.com/helioforecast}). The statistical analysis was carried out using Matlab solar wind analysis scripts OSEA. Furthermore, the exact script that was used to create the results and plots shown here (in Jupyter Notebook format) is available. See Section \ref{sec:sources} for the links to the data and scripts, and instructions for installation in the repositories. R.L.B, C.M., M.A.R., A.J.W., U.V.A., T.A. and J.H. thank the Austrian Science Fund (FWF): P31659-N27, J4160-N27, P31521-N27, P31265-N27. We would like to thank the editor and reviewers for their helpful responses, which led to improvement of this manuscript. 

%% ------------------------------------------------------------------------ %%
%% References and Citations

%%%%%%%%%%%%%%%%%%%%%%%%%%%%%%%%%%%%%%%%%%%%%%%
%
% \bibliography{<name of your .bib file>} don't specify the file extension
%
% don't specify bibliographystyle
%%%%%%%%%%%%%%%%%%%%%%%%%%%%%%%%%%%%%%%%%%%%%%%

\bibliography{l5pred}

%Reference citation instructions and examples:
%
% Please use ONLY \cite and \citeA for reference citations.
% \cite for parenthetical references
% ...as shown in recent studies (Simpson et al., 2019)
% \citeA for in-text citations
% ...Simpson et al. (2019) have shown...
%
%
%...as shown by \citeA{jskilby}.
%...as shown by \citeA{lewin76}, \citeA{carson86}, \citeA{bartoldy02}, and \citeA{rinaldi03}.
%...has been shown \cite{jskilbye}.
%...has been shown \cite{lewin76,carson86,bartoldy02,rinaldi03}.
%... \cite <i.e.>[]{lewin76,carson86,bartoldy02,rinaldi03}.
%...has been shown by \cite <e.g.,>[and others]{lewin76}.
%
% apacite uses < > for prenotes and [ ] for postnotes
% DO NOT use other cite commands (e.g., \citet, \citep, \citeyear, \nocite, \citealp, etc.).
%

\end{document}